\documentclass[12pt]{article}
\usepackage{graphicx}
\usepackage{epsf}
\newcommand{\be}[1]{\begin{equation}\label{#1}}
\newcommand{\ee}{\end{equation}}
\newcommand{\bea}[1]{\begin{eqnarray}\label{#1}}
\newcommand{\eea}{\end{eqnarray}}
\newcommand{\gl}[1]{Eq.\,(\ref{#1})}
\newcommand{\gln}[2]{Eqs.\,(\ref{#1}) and (\ref{#2})}
\newcommand{\abb}[1]{Fig.\,\ref{#1}}

\begin{document}
\vspace*{2cm}
\begin{center}
{\LARGE \bf  Does Recycling in Germinal Centers Exist?}\\
\vspace{1cm}
Michael Meyer-Hermann\\
\vspace{1cm}
Institut f\"ur Theoretische Physik, TU Dresden,
D-01062 Dresden, Germany\\
E-Mail: meyer-hermann@physik.tu-dresden.de\\
\end{center}

\vspace{2cm}
\noindent{\bf Abstract:}
A general criterion is formulated in order to decide if
recycling of B-cells exists in GC reactions.
The criterion is independent
of the selection and affinity maturation process and 
based solely on total centroblast population arguments. 
An experimental
test is proposed to verify whether the criterion is fulfilled.
\vspace*{\fill}
\eject
\newpage

\section{Introduction}

The affinity maturation process in germinal center (GC) reactions
has been well characterized in the last decade. Despite a lot
of progress concerning the morphology of GCs and the stages of the
selection process, a fundamental question remains unsolved:
Does recycling exist? {\it Recycling} means a back differentiation
of antibody presenting centrocytes (which undergo the selection
process and interact with antigen fragments or T-cells) to
centroblasts (which do not present antibodies and therefore do
not interact with antigen fragments but proliferate and mutate).

The existence of recycling in the GC has important consequences for
the structure of the affinity maturation process in GC reactions.
The centroblasts proliferate and mutate with high rates
in the environment of follicular
dendritic cells. During the GC reaction they differentiate to
antibody presenting centrocytes which may then be selected by
interaction with antigen fragments and T-cells.
If the positively
selected centrocytes recycle to (proliferating and mutating)
centroblasts the antibody-optimization process in GCs may be
compatible with random mutations of the centroblasts. 
On the contrary, one-pass GCs (without recycling)
seem to be inconsistent with an optimization process which is
based on undirected mutations. As a consequence, the search for mechanisms
which facilitate the finding of the optimal antibody-type in
GC reactions through somatic hypermutations is directly correlated
to the existence of a centrocyte recycling process.

The recycling hypothesis was first formulated in 1993 \cite{Kep93} 
and there exist several approaches to verify it.
(e.g.~\cite{Han95,Opr00,Mey01}).
Experimentally, recycling of B-cells has never been observed
directly. This may be due to a small number of recycled cells
compared to the huge total number of cells in the GC.
An indirect experimental evidence \cite{Han95} for cell recycling
was found by observing the rates of centrocyte apoptosis and 
the total cell population decrease after injection of a second antigen
(different from the antigen which originally initiated the GC reaction) 
into a GC which already was in a later phase of 
its reaction. Both, apoptosis of centrocytes as well as the
total cell population decrease were faster than in unperturbed 
GC reactions. Interestingly, the total cell population decrease
was too fast to be solely explained by centrocyte apoptosis.
This observation lead to the conclusion that a feedback
mechanism from centrocytes to centroblasts should exist. 
Recycling is a good candidate for such a mechanism.

The major conceptual problem in the theoretical 
contributions is the number of
assumptions which are necessary to describe the affinity maturation
process. The recycling hypothesis was formulated because it
seemed unlikely that a complex optimization process (involving
up to about 9 somatic hyper-mutations \cite{Kue93,Wed97}) may occur in a
one-pass GC, i.e.~without recycling: Starting from about 12000
cells which mutate in a high-dimensional (shape) space
the probability (mutation is likely to occur
stochastically \cite{Wei70,Rad98}) for at least one cell
to find the right mutant is very small. 
It would help, if the
already affinity maturated cells could reenter the proliferation
process in order to multiply and to enhance the chance to
find the optimal mutant.

Due to the origin of the recycling hypothesis, the efforts
to show its existence were based on the influence of recycling
on the maturation process. Consequently, it was found that 
the number of maturated cells becomes substantially larger
with recycling than without \cite{Opr00}. A quantitative
description showed, that 
in order to bring the affinity maturation process into
accordance with experimental observations,
the recycling probability of
positively selected centrocytes should have as high values as
80\% \cite{Mey01}.
In both models important assumptions were made
to represent the antibody-types in a shape space (affinity classes,
dimension of shape space) and to allow
a successive affinity maturation (affinity neighborhood).

Here, a new and simpler perspective is presented.
Not the influence of recycling on the maturation process is 
examined but its influence on the total centroblast population.
The centroblasts are believed not to present antibodies, so that
they do not participate at the selection process via interaction
with the antigens presented by the follicular dendritic cells. 
Therefore, 
in GCs without recycling the total centroblast population is
independent of affinity maturation and selection.
However, this statement
may depend from the system under consideration. For instance,
the centroblasts in GCs during splenic immune responses of mice
has been observed to present antibodies \cite{Cam98}.

The analysis starts from the
presupposition that recycling does not exist. With a minimal
set of assumptions (see Sec.~\ref{req})
the implications for the time course of the total centroblast
population are described in Sec.~\ref{impli}.
The dependency of the results on some characteristic
properties of various GCs is analyzed
(see Sec.~\ref{robust})
and an experimental test of the implications
is proposed (see Sec.~\ref{conclude}).


\section{Requirements}
\label{req}

\subsection{General assumptions}

In this Gedanken experiment 
the total centroblast population $B(t)$ in a 
one-pass GC (without recycling) is considered. 
{\it Total} means, that the encoded
antibody-type is ignored and all centroblasts enter
the total population with equal weight (despite any
differences in affinity to a presented antigen).

Presuming a one-pass GC no centrocytes are recycled
to centroblasts, so that the total centroblast population
becomes independent of the number of centrocytes.
Basically, only centrocytes present antibodies and thus may undergo a
selection process. As recoil effects to the total centroblast
population are excluded, this magnitude is also independent
of any properties or dynamics of the selection process.
Especially, no influence of the already achieved affinity
maturation during the optimization process exists.
Also, 
neither the number of positively selected centrocytes, 
nor the speed of the selection process,
nor the interaction with T-cells,
nor the fate of the rejected centrocytes, 
nor the production of plasma and memory cells 
have any influence on the total centroblast population.

The total centroblast population changes exclusively according
to centroblast proliferation 
and the differentiation of centroblasts to centrocytes.
Therefore, $B(t)$ follows a differential equation of the following type
\be{Ratengleichung}
\frac{dB}{dt}(t) \;=\; -f(t) B(t) \;\equiv\; p(t) B(t) - g(t) B(t)   \;,
\ee
with time $t$, centroblast proliferation rate $p(t)$, 
rate of centroblast differentiation to centrocytes $g(t)$,
and total rate function $f(t)=g(t)-p(t)$.

Cell proliferation has unquestionably to be described by
an exponential increase of the population.
Therefore, a linear proliferation term is used
at every time of the GC reaction.
In the same way the differentiation term
should be directly proportional to
the centroblast population at every moment. 
Therefore, nonlinear terms do not enter 
the total centroblast population 
dynamics \gl{Ratengleichung}.

If the centroblasts participate at the selection process as
observed in splenic immune responses in mice \cite{Cam98},
i.e.~if the centroblasts may, as the centrocytes do, undergoe 
apoptosis, Eq.~\ref{Ratengleichung} remains correct but with
a slightly modified interpretation: The growth of the number
of centroblasts is inhibited by their apoptosis. 
Therefore, the function
$p(t)$ denotes the proliferation rate of centroblasts reduced
by the centroblast apoptosis rate. This should be beared in mind
for the interpretation of the results.

Note, that no place is determined where proliferation or
differentiation occurs. Therefore, the following argument is
not restricted by the appearance or not-appearance of dark
and light zones \cite{Nos91,Cam98} during a GC reaction.


\subsection{About the rates}

It is assumed that after an initial pure proliferation phase of
some seeder cells, the total centroblast population does not
further increase. This is in accordance with measurements of the
follicle center volume, which show a peak 3 or 4 days after
immunization \cite{Liu91}.
As a consequence, the function $f(t)$ in \gl{Ratengleichung} is
positive during the whole GC reaction (started at $t=0$),
i.e.~after the first pure proliferation phase.

The proliferation rate is probably constant during the
whole GC reaction. This has not been shown experimentally in
all details but is well established by a lot of independent data
taken at different moments of the GC reaction
\cite{Liu91,Han64,Zha88}.
All these data lead to the same large proliferation rate of
\be{p}
\frac{p(t)}{\ln(2)} \,\approx\,\frac{1}{6\,h}   \,.
\ee
Nevertheless, for the following considerations the
value of the proliferation rate will not be fixed.
In order to assure generality of the argument
proliferation rates according to \gl{p} will be discussed 
as special case only. Therefore, the results of the
following analysis apply also to GCs with a
centroblast proliferation rate that varies in time.

The situation is more complicated in the case of the differentiation
rate $g(t)$. This value is not known experimentally, such that 
its absolute number as well as its time course are unclear.
From the injection of BrdUrd (5-bromo-2'-deoxyuridine)
it is possible to identify the cells which were in cell cycle
during the preceding 6 hours. 
In this way it was found in GCs with established
dark and light zones, that the centrocyte population is renewed
from centroblasts every 7 hours \cite{Liu91}. It follows for
the differentiation rate that
\be{g}
\frac{g(t)}{\ln(2)} \,>\,\frac{1}{7\,h}   \,.
\ee
Because of the lack of more detailed information,
different possibilities for the time dependence
of the centroblasts to centrocytes differentiation rate
will be considered in the following argument.

\subsection{A standard germinal center}
\label{standard}

Before analyzing the implications of this Gedanken experiment
a typical GC has to be defined as reference system.
The GC reaction is initiated by a few seeder cells in the
environment of follicular dendritic cells (FDC) 
\cite{Kue93,Liu91,Jac91,Kro87}.
These cells proliferate for about 3 days, and give rise
to about 12000 centroblasts \cite{Liu91}. At this time $t_0=0\,h$
the intrinsic GC reaction starts, including a continuation
of proliferation and, additionally, the differentiation of
centroblasts to centrocytes. The analysis made here does not
consider the first phase of pure proliferation, but deals
solely with the GC properties in the second phase.
The life time $T$ of a germinal center is about 18 days
($T=432\,h$) \cite{Liu91,Kel96}, 
in which the first phase of pure proliferation is not included.
At the end of the GC reaction, only a few cells remain in the
FDC environment \cite{Liu91,Kel96}, which is assumed to be 
about 5 cells in a standard GC. 
The sensitivity of the results on these assumptions
will be analyzed (see Sec.~\ref{robust}).


\section{Analysis of the centroblast population}
\label{impli}

\begin{figure}[ht]
\center{\hspace*{0mm} \epsfxsize=12cm \epsfysize12cm
        \epsfbox{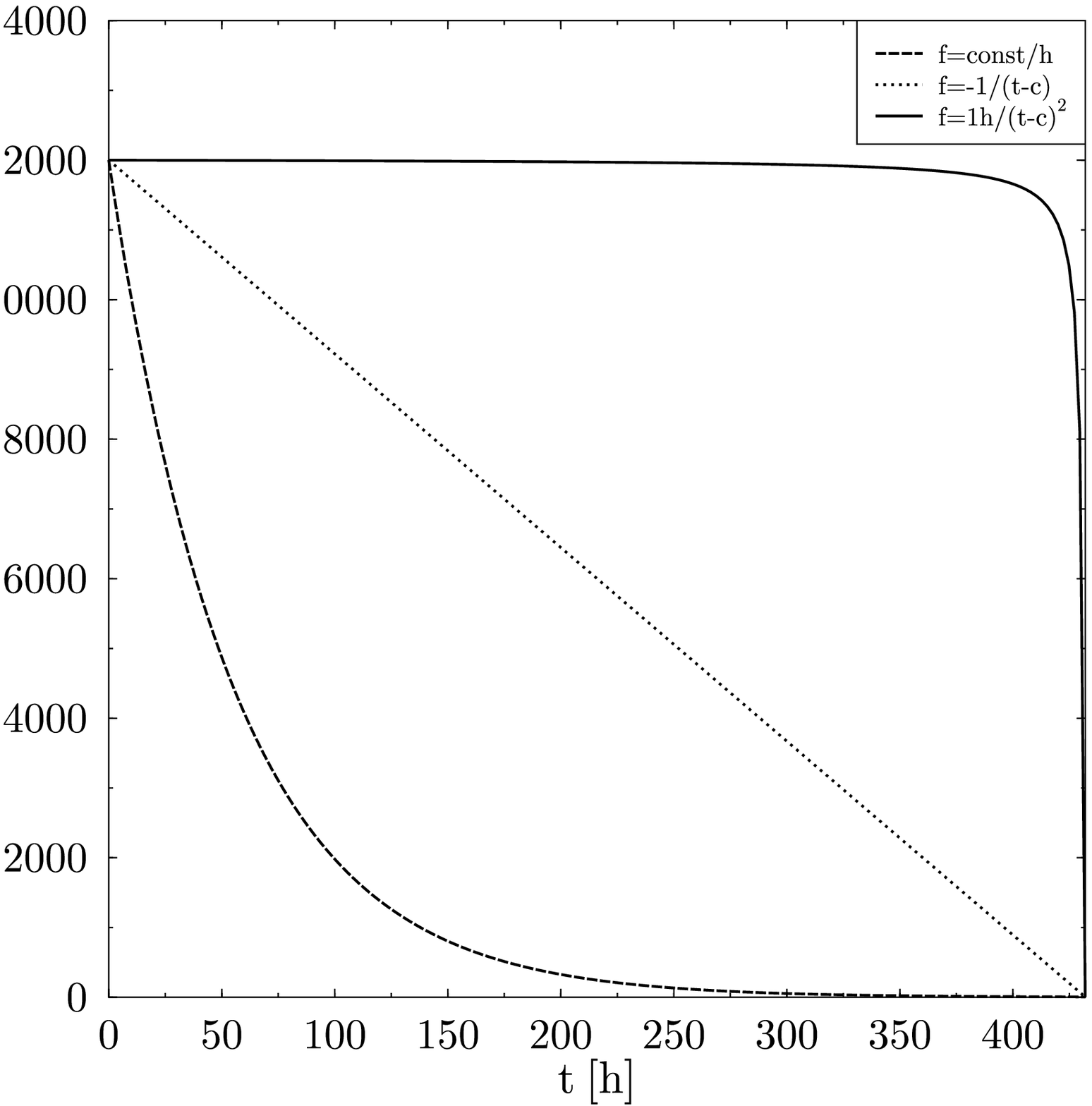}}
\caption[]{\sf Considered time courses of the centroblast population during
GC reactions. The model results remain valid for all GCs with centroblast
populations within these extreme scenarios.}
\label{bvont}
\end{figure}
\begin{figure}[ht]
\center{\hspace*{0mm} \epsfxsize=12cm \epsfysize12cm
        \epsfbox{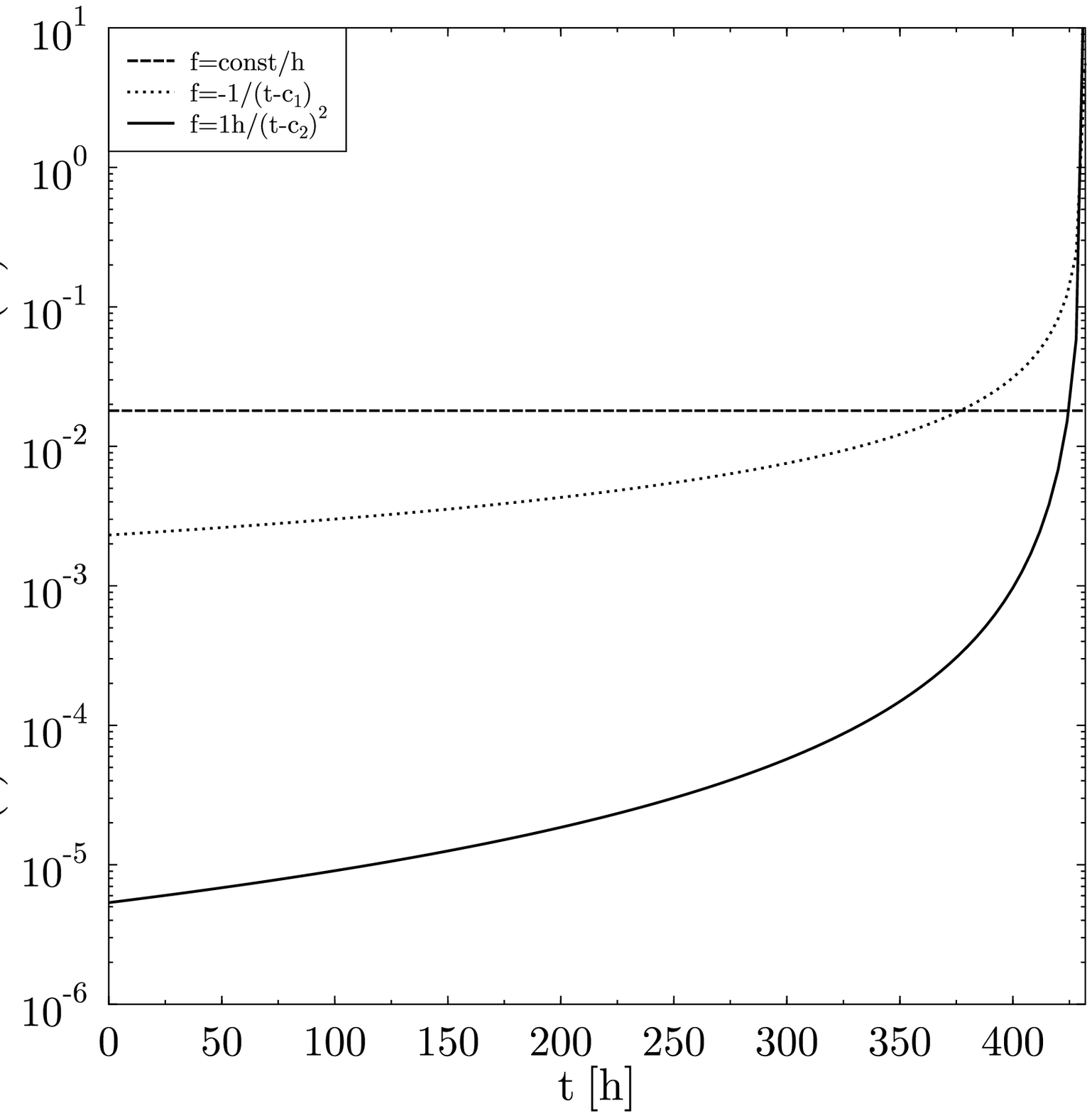}}
\caption[]{\sf The function $f(t)=g(t)-p(t)$ corresponding to the different
time courses of the centroblast population in \abb{bvont}. The GC
life time is $T=432\,h$. During the major life time of the GC the rate function
respects the upper bound $f(t)<0.018/h$ (see Eq.~\ref{condf}).}
\label{fvont}
\end{figure}
The different time courses considered for the total centroblast population
of a standard GC (defined in Sec.~\ref{standard})
are shown in \abb{bvont}. An exponential decrease corresponds
to a constant rate function (see \abb{fvont} dashed line).
A linear population decrease is achieved with the rate function
\gl{loesf} (see \abb{fvont} dotted line) which is derived
in the Appendix. Finally, using a quadratic
rate function \gl{f2} (see \abb{fvont} full line) the centroblast
population remains unchanged for the major duration of the
GC reaction and is steeply reduced at the very end.

Experimentally, the time course of the 
GC total centroblast population seems to lie 
in between the exponential and the linear 
scenario \cite{Liu91,Hol92}.
Anyhow, the scenario with a long-lasting constant centroblast
population, i.e.~ with the quadratic rate function,
is certainly a reasonable upper bound for
possible time courses of the total centroblast population
(\abb{bvont} full line).
On the other hand, a population decrease that is
faster than exponential is unlikely. 

It follows that during the major life time of GCs ($t<15\,d$) the
condition
\be{condf}
f(t) \;=\; g(t)-p(t) \;<\; 0.018/h
\ee
holds (see \abb{fvont}).
On the other hand the requirement of a monotonically decreasing
centroblast population leads to the bound $f(t)>0$, i.e.
$g(t)>p(t)$. Taking these results together it follows
\be{gcond}
p(t) \;<\; g(t) \;<\; p(t) + 0.018/h
\;,
\ee
which is a very powerful condition. This means that in an experiment
the centroblast to centrocyte differentiation rate $g(t)$ 
has to fulfill this condition during the first $15$ days of the GC reaction.
If it turns out that the above condition is not guaranteed
experimentally, then 
-- assuming the rather weak requirements made to be valid --
a GC reaction without recycling of centrocytes to centroblasts
is ruled out.

Note, that assuming the proliferation rate to adopt the value
\gl{p}, the above condition can be reformulated.
It follows a lower bound for the differentiation rate
which is in accordance with \gl{g}:
\be{gcond2}
\frac{1}{6\,h} \;<\; \frac{g(t)}{\ln(2)} \;<\; \frac{1}{5.19 \,h}
\;.
\ee
This means that if GCs without recycling exist,
a centroblast shall (in average) take not 
less than $5.19$ hours to differentiate into centrocytes.

Supposing that the linear centroblast population decrease
is the realistic scenario, 
the upper bound becomes even more powerful.
During the first 10 days the condition
\be{gcondlin}
p(t)\,<\,g(t)\,<\,p(t)+0.005/h
\ee
must hold in a GC reaction without recycling.
With the proliferation rate in \gl{p} this corresponds to
an average centroblast differentiation time of at least $5.75$ hours.
This result demonstrates, that the condition for a one-pass GC
may become stronger than in \gln{gcond}{gcond2},
depending on the characteristics of the total centroblast population
decrease.


\subsection{Variation of the GC properties}
\label{robust}

\subsubsection{The case of constant rates}

In order to check the robustness of the results \gln{gcond}{gcond2}
the most critical scenario (i.e.~with weakest conditions) 
is considered in some more detail: the scenario with constant rates.
If the proliferation and differentiation rate are constant, 
the centroblast population
model reduces to a pure exponential decrease of the population
\be{Bgconst}
B(t) \,=\, B(0)\, \exp\left((p-g)t\right)  \,,
\ee
where the constant rate $g$ is unknown.
Focusing on the population after $T=18$ days of the GC reaction,
the rate function
\be{gvonB}
f\,=\,g -p\,=\, - \frac{1}{T}\,\ln\left(\frac{B(T)}{B(0)}\right)
\ee
is found.
The resulting values for the differentiation rate are shown in
\abb{varbt} in dependence of the GC life time and for different
numbers of remaining cells at the end of the reaction.
\begin{figure}[ht]
\center{\hspace*{0mm} \epsfxsize=12cm \epsfysize12cm
        \epsfbox{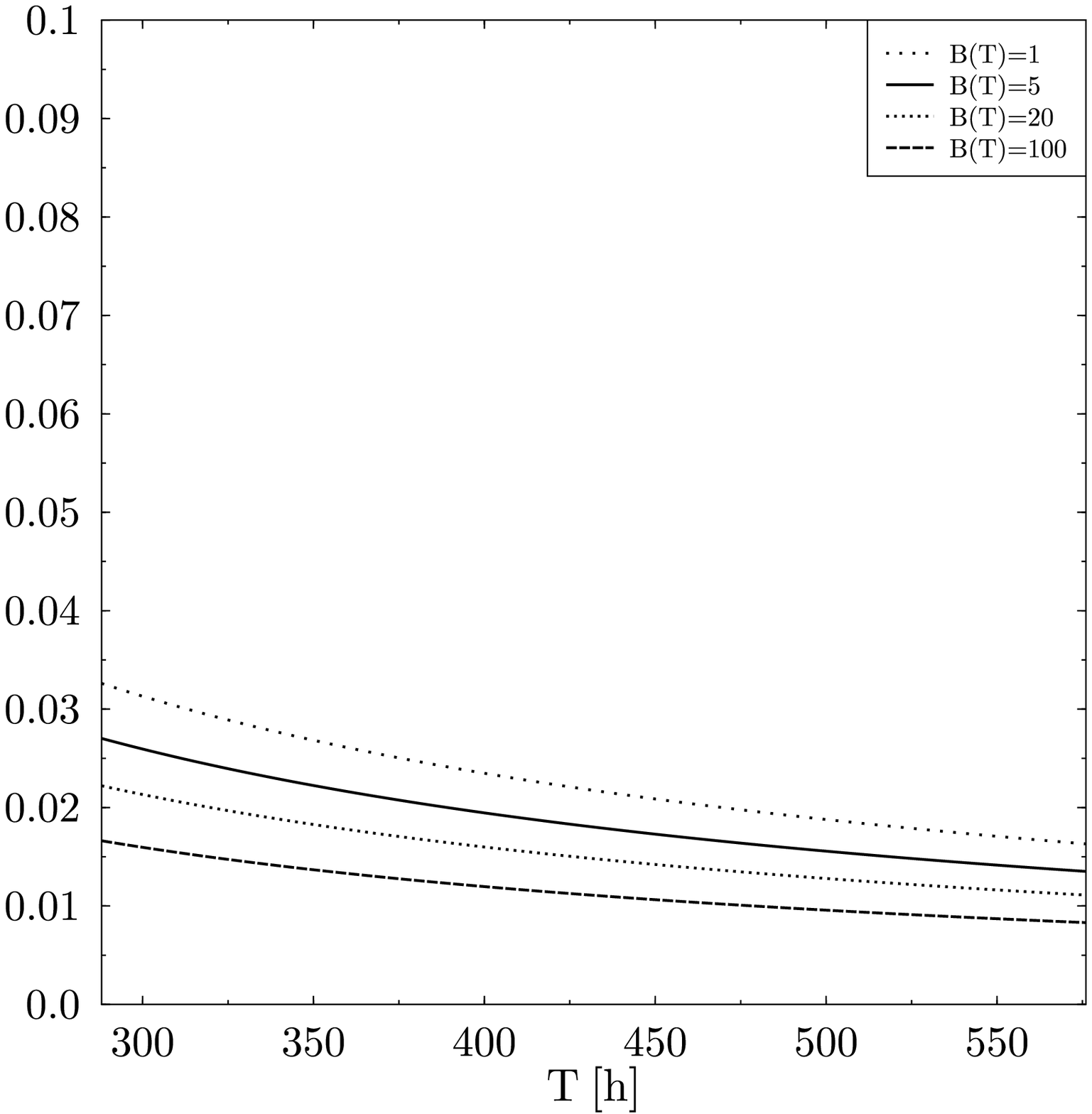}}
\caption[]{\sf The values of the (constant) rate function $f$ 
in the case of an exponentially decreasing centroblast population.
The dependence of condition Eq.~\ref{gcond} of
the GC life time and of the number of remaining 
centroblasts at the end of the GC reaction is shown to
be weak (see also Eq.~\ref{schlimmst} for the most
pessimistic case).}
\label{varbt}
\end{figure}
As expected the rate becomes larger for shorter GC life times $T$.
Even supposing a rather short life time of 12 days 
and further supposing that
only one cell remains at the end of the GC reaction, 
one gets an upper bound of 
\be{schlimmst}
p(t) \;<\; g(t) \;<\; p(t) + 0.033/h
\ee
for possible values of the differentiation rate. Incorporating 
the proliferation rate \gl{p} it follows
\be{schlimm2}
\frac{1}{6\,h} \;<\; \frac{g(t)}{\ln(2)} \;<\; \frac{1}{4.7 \,h}
\,.
\ee
As a consequence, the statement
that in one-pass GCs (without recycling) the centroblast to centrocyte 
differentiation rate 
should respect the above bounds \gln{gcond}{gcond2} is not
altered dramatically by variation of the GC properties.

The initial number of centroblasts $B(0)$ was not varied so far. 
Note that
only the ratio $B(T)/B(0)$ enters \gl{gvonB}, so that a variation of
the initial number of centroblast is equivalent to the variation of
the final number of remaining cells $B(T)$. Therefore, the
constraints \gln{schlimmst}{schlimm2} are also valid
with e.g.~60000 initial and 5 remaining cells. Lower numbers of
initial cells allow even stronger bounds than the above ones.

\subsubsection{The case of linear centroblast decrease}

\begin{figure}[ht]
\center{\hspace*{0mm} \epsfxsize=12cm \epsfysize12cm
        \epsfbox{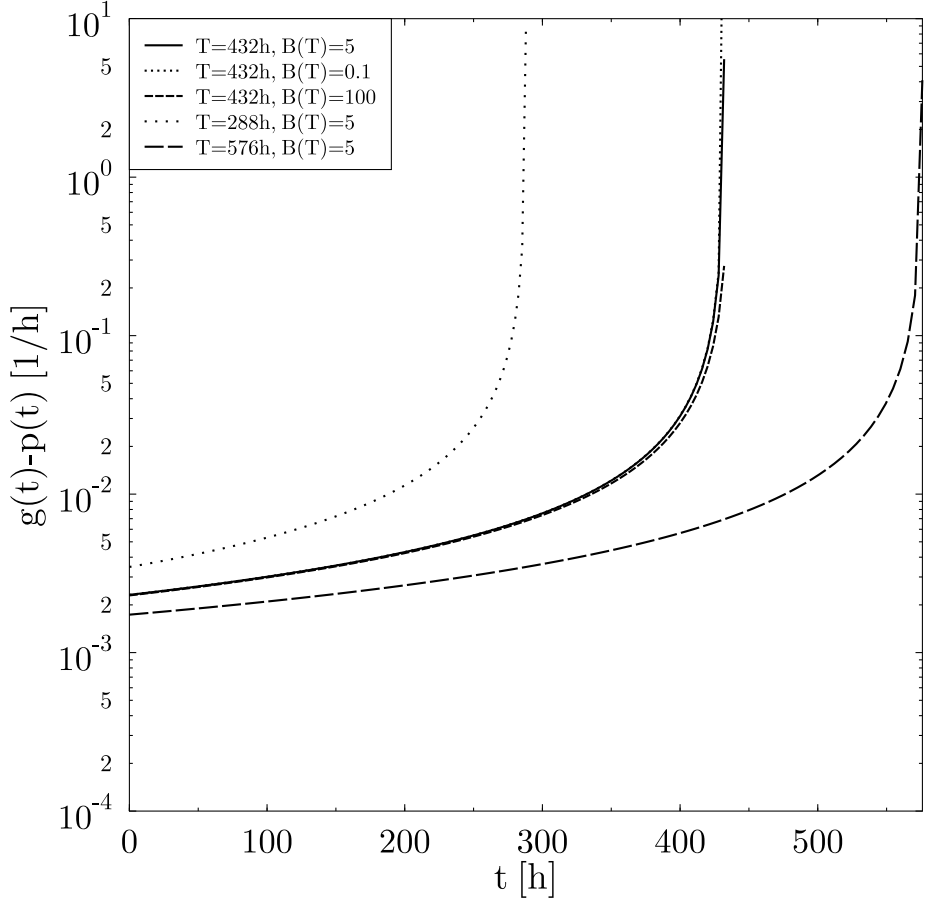}}
\caption[]{\sf The time course of the rate function $f=g-p$ 
(corresponding to a linear centroblast population decrease)
for different GC life times $T$ and for different numbers of remaining 
centroblasts $B(T)$ at the end of the GC reaction.
Condition Eq.~\ref{gcondlin} turns out not to depend
on the number of remaining centroblasts.
The rate function is sensitive to the total GC life time
in the late phase of the reaction. However, condition
Eq.~\ref{gcond} remains valid for practically the whole
GC life time.}
\label{varfvont}
\end{figure}
In the case of a linear centroblast population decrease 
(see \abb{bvont} pointed line)
the corresponding rate function $f(t)$ (see \gl{loesf}) 
is analyzed for various GC life times $T$ and numbers of remaining
blasts at the end of the reaction $B(T)$. The result is shown in
\abb{varfvont}. Condition \gl{gcondlin} is not affected at all by
a variation of the number of remaining cells at the end of the GC
reaction in a range of three orders of magnitude (the values chosen
for $B(T)$ are $0.1$, $5$, and $100$). 
The three corresponding curves are practically not distinguishable (see
\abb{varfvont}). 

The situation is different for the variation
of the GC life time $T$. Here, the period in which condition \gl{gcondlin}
remains valid is prolonged for long living GCs and shortened for
short living GCs (see \abb{varfvont} dashed and pointed line, respectively). 
For a very short life time of 12 days (plus
3 days of pure proliferation) the condition remains valid for 4 days only
(7 days after immunization). However, condition \gl{gcond} remains valid even 
in this case for 11 days, so for approximately the whole GC life time.


\section{Conclusions}
\label{conclude}

The above analysis is summarized with the statement that 
in a one-pass GC (without recycling) the centroblast to
centrocyte differentiation rate should respect the strong
condition \gl{gcond} $p(t)<g(t)<p(t)+0.018/h$ during the
first 15 days of the GC reaction (with standard properties
which are defined in \ref{standard}).
For a standard proliferation rate \gl{p} of $\ln(2)/6h$
this translates into an average time for
centroblasts to differentiate into centrocytes which
cannot be substantially shorter than $5$ hours. 
If this condition is not fulfilled, recycling has
necessarily to be present in GC reactions.
Note that, the other way round, if the average differentiation
of centroblasts to centrocytes takes more than 5 hours, recycling
is not necessarily absent.

This statement is based essentially on two (weak) assumptions: 
Firstly, linear population dynamics (see \gl{Ratengleichung}) for the
total centroblast population in GCs. Secondly, a monotonic decrease
of the total centroblast population. 
The conclusions apply universally to GCs with various 
characteristics due to
the independence of all speculations on shape spaces, 
selection processes, and affinity maturation.
However, one should be aware that still some assumptions
may be questioned.
Remind, for instance, 
that for GCs with selectable centroblasts (that may
undergoe apoptosis) the rate $p$ has to be interpreted as
the rate of centroblast proliferation reduced by the 
centroblast apoptosis rate. 

In order to decide if recycling does exist in GC reactions
one has to determine the centroblast to centrocyte differentiation
rate experimentally {\it at least} at one representative moment of the
GC reaction. The moment has to be chosen such that on one hand
the GC reaction is already well established. On the other hand,
the later the measurement is done the less is known on the
proliferation rate. 
In addition, one approaches the regime where
the above condition looses its validity (see \abb{fvont}).
Therefore, an optimal time will be about 8 days after
immunization.
However, if one expects the centroblast proliferation rate to vary
substantially during this central phase of GC reaction, 
the standard values for $p$ cannot be used. 
In this case, one would have to determine
not only the centroblast differentiation rate but also
the centroblast proliferation rate.

Technically, the centroblast differentiation rate 
may be measured by labeling studies of
B-cells in the GC which are in cell cycle as it was done
in \cite{Cam98,Liu91}.\\
Alternatively, a systematic and time resolved analysis
of the total centroblast population 
at different moments of the GC reaction may improve the
criterion (see e.g.~\gl{gcondlin}) because one could distinguish which
of the cell population scenarios presented in \abb{bvont}
corresponds to real GC reactions. 
This may be done by
the measurement of the GC volume at several times of the
GC reaction as in \cite{Liu91,Hol92,Vin00}.


\section*{Appendix}

The differential equation \gl{Ratengleichung} is formally
solved by
\be{loes}
B(t) \;=\; B(t_0=0\,h)\, \exp\left(-\int_0^t dx\,f(x)\right)
\;.
\ee
In the following, the border conditions 
(life time of GC $T$, and centroblast population at $t_0=0\,h$
and $T$) are incorporated into the rate function $f$.

To achieve a linear decrease of the centroblast population, 
the second time derivative of the population is required to
vanish. From \gl{Ratengleichung} and using \gl{loes} it follows
\bea{B_tt}
\frac{d^2B}{dt^2}
&=&
-\frac{df}{dt}B -f\frac{dB}{dt} \nonumber\\
&=&
-\frac{df}{dt} B(0) \, \exp\left(-\int_0^t dx\,f(x)\right)
+f B(0) f \, \exp\left(-\int_0^t dx\,f(x)\right)
\;.
\eea
This expression vanishes for
\be{dglf}
\frac{df}{dt}=f^2
\;,
\ee
which is solved by
\be{loesf}
f(t)\;=\; -\frac{1}{t-c_1} \;.
\ee
The integration constant $c_1$ is fixed by the condition that the
final population at time $T$ is given by $B(T)$. With \gl{loes}
this leads to
\be{c1}
c_1\;=\;\frac{T}{1-\frac{B(T)}{B(0)}}
\;,
\ee
which is a positive constant for $B(T)<B(0)$.

In the case of a quadratic rate function 
\be{f2}
f(t) \;=\; \frac{(1\,h)}{(t-c_2)^2}
\ee
an analogous calculation which incorporates the final
centroblast populations leads to
\be{c2}
c_2\;=\;\frac{T}{2} + 
\sqrt{\frac{T^2}{4}-\frac{(1\,h)\,T}{\ln\left(\frac{B(T)}{B(0)}\right)}}
\;.
\ee
Other solutions are formally possible but do not lead to biologically
reasonable centroblast populations.


\vspace*{5mm}
\subsection*{Acknowledgments}
I thank Andreas Deutsch and Michal Or-Guil
for intense discussions and valuable comments.

\end{document}